\documentclass{article}

\usepackage{amsmath}
\usepackage{graphicx}
\usepackage{bm}
\usepackage{psfrag}

\newcommand{\dif}{\,\mathrm{d}}
\newcommand{\ldirac}[1]{\langle #1 |}
\newcommand{\rdirac}[1]{|#1\rangle}

\begin{document}
\parbox{13 cm}
{
\begin{flushleft}
\vspace* {1.2 cm}
{\Large\bf {Atoms near magnetodielectric bodies: van-der-Waals energy and Casimir-Polder force}
}\\
\vskip 1truecm
{\large\bf
{
S. Y. Buhmann$^{1,2}$), Ho Trung Dung$^{3}$), T. Kampf$^1$), L. Kn\"{o}ll$^1$),
and D.-G. Welsch$^1$)
}
}\\
\vskip 5truemm
{
$^1$) Theoretisch-Physikalisches Institut,
Friedrich-Schiller-Universit\"{a}t Jena, Max-Wien-Platz 1,
07743 Jena, Germany\\
$^2$) Electronic address: \texttt{s.buhmann@tpi.uni-jena.de}\\
$^3$) Institute of Physics, National Center for Natural Sciences and
Technology, 1 Mac Dinh Chi Street, District 1, Ho Chi Minh City, Vietnam
}

\end{flushleft}
}
\vskip 0.5truecm
{\bf Abstract:\\}

{
\noindent
Based on macroscopic QED in linear, causal media, we present a consistent theory for the Casimir-Polder force acting on
an atom positioned near dispersing and absorbing magnetodielectric bodies. The perturbative result for the van-der-Waals
energy is shown to exhibit interesting new features in the presence of magnetodielectric bodies.
To go beyond perturbation theory, we start with the center-of-mass equation of motion and derive a dynamical expression
for the Casimir-Polder force acting on an atom prepared in an arbitrary electronic state. For a non-driven atom in
the weak coupling regime, the force as a function of time is shown to be a superposition of force components that are
related to the electronic density matrix elements at chosen time. These force components depend on the 
position-dependent polarizability of the atom that correctly accounts for the body-induced level shifts and broadenings.
}

\section{Introduction}
\label{sec1}

Being a result of the vacuum fluctuations of the electromagnetic field, Casimir-Polder (CP) forces are experienced by
any atomic system in the presence of magnetodielectric bodies. They play an important role in physical chemistry
\cite{Chesters73}, and they hold the key to potential applications in micro- and nanotechnology such as the construction
of atomic-force microscopes \cite{Binnig86} or reflective atom-optical elements \cite{Shimizu02}.

On short time scales and for weak atom-field coupling, the CP force is commonly
derived from the van-der-Waals (vdW) energy calculated by means of time-independent perturbation
theory \cite{Casimir48,Milloni92,Buhmann03} or linear response theory \cite{McLachlan63,WylieSipe,Henkel02}.
Following the former approach, but using a quantization scheme for the electromagnetic field in the presence of
dispersing and absorbing magnetodielectric bodies, we derive a general expression for the vdW energy of an atom prepared
in an energy eigenstate, where we focus on the influence of the magnetic properties.

The failure of perturbation theory is evident, because it cannot account for the internal dynamics present for an atom
initially prepared in an excited state, and because the leading-order atomic polarizability, which essentially
determines the CP force acting on an atom in the ground state, does not incorporate the body-induced energy shifts and
broadenings, which can become drastic for small atom-surface separations. To overcome this deficiency, we present a
dynamical treatment by basing the calculations on the well-known Lorentz force that governs the atomic center-of-mass
equation of motion. The CP force is then obtained by taking the average of the Lorentz force with respect to the
electronic quantum state of the atom and the electromagnetic vacuum, resulting in a dynamical expression that is valid
for both strong and weak atom-field coupling. For the case of weak atom-field coupling, this expression is further
evaluated with the aid of the Markov approximation.


\section{Basic equations}
\label{sec2}

Let us consider a neutral atomic system (e.g. an atom or a molecule) interacting with the electromagnetic field in the
presence of linear, causal, magnetodielectric media and begin with the Hamiltonian
\begin{equation}
\label{eq1}
     \hat{H} =  \hat{H}_{\rm A} + \hat{H}_{\rm F} + \hat{H}_{\rm AF}.
\end{equation}
According to the multipolar coupling scheme \cite{Craig84},
\begin{equation}
\label{eq2}
      \hat{H}_{\rm A} =
      \sum_{\alpha}
      \frac{\hat{{\bf p}}_{\alpha}{\!^2}}{2m_{\alpha}}
      + \frac{1}{2\varepsilon_0} \int{\rm d}^3r\,
      \hat{{\bf P}}_{\rm A}^2 ({\bf r})
\end{equation}
is the Hamiltonian for the atomic system consisting of particles $\alpha$ with charges $q_\alpha$, masses $m_\alpha$,
positions $\hat{\bf r}_{\alpha}$, and canonically conjugated momenta $\hat{\bf p}_{\alpha}$, where
\begin{equation}
\label{eq3}
     \hat{{\bf P}}_{\rm A}({\bf r}) =
     \sum_\alpha q_\alpha
     {\hat{\bar{\bf r}}}_\alpha
     \int _0^1 {\rm d}\lambda
     \,\delta({\bf r} - \hat{{\bf r}}_{\rm A}
     - \lambda{\hat{\bar{\bf r}}}_\alpha )
\end{equation}
is the atomic polarization relative to the center of mass
\begin{equation}
\label{eq4}
       \hat{{\bf r}}_{\rm A} = \sum_\alpha \frac{m_\alpha}{m_{\rm A}}
       \,\hat{{\bf r}}_\alpha
\end{equation}
($m_{\rm A}$ $\!=$ $\!\sum_\alpha m_\alpha$),
\begin{equation}
\label{eq5}
      {\hat{\bar{\bf r}}}_\alpha=\hat{\bf r}_\alpha
      -\hat{{\bf r}}_{\rm A}
\end{equation}
denoting shifted particle coordinates. The Hamiltonian $\hat{H}_\mathrm{F}$ characterizing the medium-assisted
electromagnetic field is given by \cite{Knoll01,Ho03}
\begin{equation}
\label{eq6}
      \hat{H}_{\rm F} \equiv
      \sum_{\lambda=e,m}\int{\rm d}^3 {r}
      \int_0^{\infty}{\rm d}\omega\,\hbar\omega\,
      \hat{\bf f}_\lambda{\!^\dagger}({\bf r},\omega)
      \hat{\bf f}_\lambda({\bf r},\omega),
\end{equation}
where the bosonic fields $\hat{\bf f}_\lambda({\bf r},\omega)$ [and $\hat{\bf f}_\lambda^\dagger({\bf r},\omega)$],
\begin{equation}
\label{eq7}
      \left[\hat{f}_{\lambda i}({\bf r},\omega),
      \hat{f}^\dagger_{\lambda' i'}({\bf r'},\omega')\right]
      =\delta_{\lambda\lambda'}
      \delta_{ii'}\delta({\bf r}-{\bf r}')\delta(\omega-\omega'),
\quad
      \left[\hat{f}_{\lambda i}({\bf r},\omega),
      \hat{f}_{\lambda' i'}({\bf r'},\omega')\right]
      =0,
\end{equation}
play the role of the dynamical variables of the electromagnetic field plus the medium, where $\lambda$ $\!=$ $\!e$ and
$\lambda$ $\!=$ $\!m$, respectively, refer to the electric and magnetic properties of the medium. Finally, the
multipolar-coupling Hamiltonian describing the interaction between the atomic system and the medium-assisted
electromagnetic field in electric dipole approximation reads
\begin{equation}
\label{eq8}
      \hat{H}_{\rm AF}=
      - \hat{{\bf d}}
      \hat{{\bf E}}(\hat{{\bf r}}_{\rm A})
      +\frac{1}{2m_{\rm A}}
        \bigl[\hat{\bf p}_{\rm A},\hat{\bf d}\!\times\!
        \hat{{\bf B}}(\hat{\bf r}_{\rm A})\bigr]_+
\end{equation}
[$\hat{{\bf E}}(\hat{{\bf r}})$, electric field; $\hat{{\bf B}}(\hat{{\bf r}})$, induction field;
$[\hat{\bf a},\hat{\bf b}]_+=\hat{\bf a}\hat{\bf b}+\hat{\bf b}\hat{\bf a}$], with
\begin{equation}
\label{eq9}
      \hat{\bf d}
      = \sum_{\alpha}q_{\alpha}\hat{\bf r}_{\alpha}
      = \sum_{\alpha}q_{\alpha}{\hat{\bar{\bf r}}}_{\alpha}
\end{equation}
being the electric dipole moment of the atomic system. Note that the second term on the right-hand side of
Eq.~(\ref{eq8}) describes the R\"{o}ntgen interaction due to the translational motion of the center of mass
\cite{Craig84},
\begin{equation}
\label{eq10}
 \hat{\bf p}_{\rm A}=\sum_\alpha \hat{\bf p}_\alpha
\end{equation}
denoting the total momentum of the atomic system.

The medium-assisted electric field [which in the multipolar coupling scheme has the physical meaning of a displacement
field with respect to the atomic polarization (\ref{eq3})] and the induction field can be related to the fundamental
bosonic fields via \cite{Knoll01,Ho03}
\begin{gather}
\label{eq11}
      \hat{{\bf E}}({\bf r})=\int_0^\infty {\rm d} \omega\,
      \underline{\hat{\bf E}}({\bf r},\omega)
      + {\rm H.c.},\\
\label{eq12}
      \hat{{\bf B}}({\bf r})
      = \int_0^\infty {\rm d} \omega\,
      (i\omega)^{-1}
      \bm{\nabla}\times \underline{\hat{\bf E}}({\bf r},\omega)
      + {\rm H.c.},\\
\label{eq13}
     \underline{\hat{\bf E}}({\bf r},\omega)=
      \sum_{\lambda=e,m}\int {\rm d}^3r'\,
      \bm{G}_\lambda({\bf r},{\bf r}',\omega)
      \hat{\bf f}_\lambda({\bf r}',\omega).
\end{gather}
The c-number tensors
\begin{gather}
\label{eq14}
     \bm{G}_e({\bf r},{\bf r}',\omega) = i\,\frac{\omega^2}{c^2}
     \sqrt{\frac{\hbar}{\pi\varepsilon_0}\,
     {\rm Im}\,\varepsilon({\bf r}',\omega)}\,
     \bm{G}({\bf r},{\bf r}',\omega),\\
\label{eq15}
     \bm{G}_m({\bf r},{\bf r}',\omega)
     = -i\,\frac{\omega}{c}
     \sqrt{-\frac{\hbar}{\pi\varepsilon_0}\,
     {\rm Im}\,\kappa({\bf r}',\omega)}
     \left[\bm{G}({\bf r},{\bf r}',\omega)\times
     \overleftarrow{\bm{\nabla}}_{\!\mathbf{r}'}\right]\!,
\end{gather}
where
$[\bm{G}({\bf r},{\bf r}',\omega)\times\overleftarrow{\bm{\nabla}}_{\!\mathbf{r}'}\bigr]_{ij}$
$\!=$ $\!\epsilon_{jkl}\partial'_l G_{ik}({\bf r},{\bf r}',\omega)$,
$\kappa(\mathbf{r},\omega)$ $\!=$ $\!\mu^{-1}(\mathbf{r},\omega)$, are given in terms of the (classical) Green tensor,
which is defined by the differential equation
\begin{equation}
\label{eq16}
      \left[
      \bm{\nabla}\times\kappa({\bf r},\omega)\bm{\nabla}\times
      -\frac{\omega^2}{c^2}\,\varepsilon({\bf r},\omega)
      \right]
      \bm{G}({\bf r},{\bf r}',\omega)
      = \bm{\delta}({\bf r}-{\bf r}')
\end{equation}
together with the boundary condition at infinity. Note that the (relative) permittivity $\varepsilon({\bf r},\omega)$
and permeability $\mu({\bf r},\omega)$ of the (inhomogeneous) medium are complex functions of frequency, whose real and
imaginary parts satisfy the Kramers-Kronig relations. The Green tensor has the following useful properties
\cite{Knoll01},
\begin{equation}
\label{eq17}
     \bm{G}^{\ast}({\bf r},{\bf r}',\omega)
     =\bm{G}({\bf r},{\bf r}',-\omega^{\ast}),
\end{equation}
\begin{equation}
\label{eq18}
     \bm{G}({\bf r},{\bf r}',\omega)
     =\bm{G}^\top({\bf r}',{\bf r},\omega),
\end{equation}
\begin{eqnarray}
\label{eq19}
\lefteqn{
    \int \!{\rm d}^3 s\, \Bigl\{
    {\rm Im}\,\kappa({\bf s},\omega)
    \left[
    \bm{G}({\bf r},{\bf s},\omega)
    \times\overleftarrow{{\bm{\nabla}}}_{\!\bf s}
    \right]
    \left[
    {\bm{\nabla}}_{\!{\bf s}}\times
           \bm{G}^\ast({\bf s},{\bf r}',\omega) \right]
}
\nonumber\\&&\hspace{-2ex}
    + \,\frac{\omega^2}{c^2}\, {\rm Im}\,\varepsilon({\bf s},\omega)
      \,\bm{G}({\bf r},{\bf s},\omega)
      \bm{G}^\ast({\bf s},{\bf r}',\omega)
      \Bigr\}
      = {\rm Im}\,\bm{G}({\bf r},{\bf r}',\omega),
\nonumber\\&&
\end{eqnarray}
where the last equation can be combined with the definitions (\ref{eq14}) and (\ref{eq15}) to yield the useful identity
\begin{equation}
\label{eq20}
 \sum_{\lambda=e,m}\int {\rm d}^3 s\,
 G_{\lambda ii'}({\bf r},{\bf s},\omega)
 G^\ast_{\lambda ji'}({\bf r}',{\bf s},\omega)
 =\frac{\hbar\mu_0}{\pi}\omega^2{\rm Im}\,G_{ij}({\bf r},{\bf r}',\omega).
\hspace{10ex}
\end{equation}


\section{The van-der-Waals energy}
\label{sec3}

Let us consider an atomic system at rest ($\hat{\mathbf{r}}_\mathrm{A}$ $\!\mapsto$ $\!\mathbf{r}_\mathrm{A}$) which is
prepared in an energy eigenstate $|l\rangle$, i.e., an eigenstate of the Hamiltonian $\hat{H}_\mathrm{A}$
[Eq.~(\ref{eq2})] written in the form
\begin{equation}
\label{eq22}
      \hat{H}_{\rm A}
      =\sum_n E_n |n\rangle\langle n|,
\end{equation}
and calculate, within the frame of Schr\"{o}dinger's perturbation theory, the leading-order energy shift
\begin{equation}
\label{eq23}
     \Delta E_l = \Delta E_l^{(0)} + \Delta E_l^{(1)}({{\bf r}_{\rm A}})
\end{equation}
of the state $|l\rangle|\{0\}\rangle$, where $|\{0\}\rangle$ denotes the ground state of the fundamental
fields $\hat{\bf f}_\lambda({\bf r},\omega)$. According to Casimir's and Polder's pioneering concept
\cite{Casimir48}, the position-dependent part $\Delta E_l^{(1)}({{\bf r}_{\rm A}})$ of this energy shift
can be interpreted as a potential energy
\begin{equation}
\label{eq24}
     U_l({{\bf r}_{\rm A}}) =
     \Delta E_l^{(1)}({{\bf r}_{\rm A}}),
\end{equation}
commonly called vdW energy, from which the CP force acting on the atom in the state $|l\rangle$ can be derived according
to
\begin{equation}
\label{eq25}
\mathbf{F}_l(\mathbf{r}_\mathrm{A})
 = - \bm{\nabla}_{\!\!\mathrm{A}}
 U_l(\mathbf{r}_\mathrm{A})
\end{equation}
($\bm{\nabla}_{\!\!\mathrm{A}}$ $\!\equiv$ $\!\bm{\nabla}_{\!\mathbf{r}_\mathrm{A}}$). In this approach to the problem,
the second term on the right-hand side of Eq.~(\ref{eq8}) can be disregarded, so that the interaction Hamiltonian
that gives rise to the energy shift reduces to
\begin{equation}
\label{eq21}
      \hat{H}_{\rm AF}=
      - \hat{{\bf d}}
      \hat{{\bf E}}(\hat{{\bf r}}_{\rm A}).
\end{equation}


\subsection{General result}
\label{sec3.1}

The bilinear form of the atom-field coupling Hamiltonian (\ref{eq21}) implies that the leading-order energy shift is
given by the second-order perturbative correction
\begin{equation}
\label{eq26}
\Delta E_l = -\frac{1}{\hbar}
\sum_k\sum_{\lambda=e,m}\mathcal{P}
      \int_0^{\infty}
      \frac{\dif\omega}{\omega_{kl}+\omega}\int\dif^3{r}\,
      \big|\ldirac{l}\ldirac{\{0\}}
              -\hat{\bf d}\hat{\bf E}({\bf r}_{\rm A})
             \rdirac{\{{\bf 1}_\lambda({\bf r},\omega)\}}
             \rdirac{k}\big|^2
\end{equation}
[$\mathcal{P}$, principal part; $\omega_{kl}$ $\!\equiv$ $\!(E_k-E_l)/\hbar$;
$|\{{\bf 1}_\lambda({\bf r},\omega)\}\rangle$ $\!\equiv$
$\!\hat{\bf f}^{\dagger}_\lambda({\bf r},\omega)|\{0\}\rangle$]. By recalling the definition (\ref{eq11}) together with
(\ref{eq13}), and making use of the commutations relations (\ref{eq7}) as well as the identity (\ref{eq20}),
it is straightforward exercise to show that
\begin{equation}
\label{eq27}
     \Delta E_l=
     - \frac{\mu_0}{\pi}\sum_k
     \mathcal{P}\int_0^{\infty}\!\!\dif\omega\,
     \frac{\omega^2}{\omega_{kl}+\omega}
     \,{\bf d}_{lk}
     {\rm Im}\,\bm{G} ({{\bf r}_{\rm A}},{{\bf r}_{\rm A}},\omega)
     {\bf d}_{kl}
\end{equation}
(${\bf d}_{lk} = \langle l|\hat{\bf d}|k\rangle$). In order to extract the position-dependent part of the energy shift
in accordance with Eq.~(\ref{eq23}), we note that the atomic system should be located in a free-space region, where the
Green tensor can be decomposed into the (translationally invariant) vacuum Green tensor $\bm{G}^{(0)}$ and the
scattering Green tensor $\bm{G}^{(1)}$ that accounts for the presence of magnetodielectric bodies,
\begin{equation}
\label{eq28}
     \bm{G}({\bf r},{\bf r}',\omega)
     = \bm{G}^{(0)}({\bf r},{\bf r}',\omega)
     + \bm{G}^{(1)}({\bf r},{\bf r}',\omega).
\end{equation}
The vdW energy (\ref{eq24}) can thus be obtained by making the replacement
[$\bm{G}(\mathbf{r}_\mathrm{A},\mathbf{r}_\mathrm{A},\omega)$ $\!\mapsto$
$\!\bm{G}^{(1)}(\mathbf{r}_\mathrm{A},\mathbf{r}_\mathrm{A},\omega)$]
in Eq.~(\ref{eq27}). The result can be simplifyed by exploiting the property (\ref{eq17}) and the well-known asymptotic
properties of the Green tensor for large frequencies (cf. Ref.~\cite{Ho03}) to transform the integral along the real
frequency axis into an integral along the imaginary frequency axis via contour-integral techniques, leading to
\begin{equation}
\label{eq29}
U_l({{\bf r}_{\rm A}})=U_l^{\rm or}({{\bf r}_{\rm A}})+U_l^{\rm r}({{\bf r}_{\rm A}}),
\end{equation}
where
\begin{equation}
\label{eq30}
     U_l^{\rm or}({{\bf r}_{\rm A}})
     = \frac{\mu_0}{\pi}
     \sum_k
     \!\int_0^{\infty}\!\!\! \dif u\,
     \frac{\omega_{kl} u^2}{\omega_{kl}^2 + u^2}
     \,{\bf d}_{lk} \bm{G}^{(1)}({{\bf r}_{\rm A}},
     {{\bf r}_{\rm A}},iu){\bf d}_{kl}
\end{equation}
is the off-resonant part of the vdW potential, and
\begin{equation}
\label{eq31}
U_l^{\rm r}({{\bf r}_{\rm A}})=
     -\mu_0
     \sum_k \Theta(\omega_{lk})
     \omega_{lk}^2\,
     {\bf d}_{lk}\, {\rm Re}\,\bm{G}^{(1)}({{\bf r}_{\rm A}},
     {{\bf r}_{\rm A}},\omega_{lk})
     {\bf d}_{kl}
\end{equation}
[$\Theta(z)$, unit step function] is the resonant part due to the contribution from the residua at the poles at
$\omega=\pm\omega_{lk}$ for $\omega_{lk}$ $\!>$ $\!0$. Introducing the (lowest-order) atomic polarizability
\begin{equation}
\label{eq32}
     \bm{\alpha}_l^{(0)}(\omega)=
     \lim_{\epsilon\to 0}
     \frac{2}{\hbar}\sum_k \frac{\omega_{kl}}{\omega_{kl}^2-\omega^2
     - i\omega\epsilon}
     \,{\bf d}_{lk}\otimes{\bf d}_{kl},
\end{equation}
we may rewrite Eq.~(\ref{eq30}) in the more compact form
\begin{equation}
\label{eq33}
     U_l^{\rm or}({\bf r}_{\rm A})
     = \frac{\hbar\mu_0}{2\pi}
     \int_0^{\infty} \dif u \,u^2 {\rm Tr}\bigl[\bm{\alpha}_l^{(0)}(iu)
     \,\bm{G}^{(1)}({\bf r}_{\rm A},{\bf r}_{\rm A},iu)
     \bigr].
\end{equation}
Finally, we note that for an atomic system in a spherically symmetric state we have
\begin{equation}
\label{eq34}
     \bm{\alpha}_l^{(0)}(\omega) = \alpha_l^{(0)}(\omega)\bm{I}=
     \lim_{\epsilon\to 0}
     \frac{2}{3\hbar}\sum_k
     \frac{\omega_{kl}}{\omega_{kl}^2-\omega^2
     - i\omega\epsilon}
     \,|{\bf d}_{lk}|^2\bm{I}
\end{equation}
($\bm{I}$, unit tensor) and thus Eqs.~(\ref{eq33}) and (\ref{eq31}) simplify to
\begin{gather}
\label{eq35}
     U_l^{\rm or}({\bf r}_{\rm A})
     = \frac{\hbar\mu_0}{2\pi}
     \int_0^{\infty} \dif u \,u^2 \alpha_l^{(0)}(iu)
     \,{\rm Tr}\,
     \bm{G}^{(1)}({\bf r}_{\rm A},{\bf r}_{\rm A},iu),\\
\label{eq36}
U_l^{\rm r}({\bf r}_{\rm A})=
-\frac{\mu_0}{3}\sum_k\Theta(\omega_{lk})
     \omega_{lk}^2
     |{\bf d}_{lk}|^2
     \,{\rm Tr}\,\bigl[{\rm Re}\,\bm{G}^{(1)}({{\bf r}_{\rm A}}
     {{\bf r}_{\rm A}},\omega_{lk})\bigr].
\end{gather}

Equation (\ref{eq29}) together with Eqs.~(\ref{eq31}) and (\ref{eq33}) is an extension of previous results
\cite{Buhmann03} to the case of arbitrary causal magnetodielectric bodies. It is worth noting that Eq.~(\ref{eq29}) also
applies to left-handed material \cite{Veselago68}, for which standard quantization concepts run into difficulties. The
calculations presented here can be regarded as the natural foundation for similar results obtained on the basis of
(semi-classical) linear response theory for dielectric bodies \cite{WylieSipe}.

\begin{figure}[htb]
\begin{center}
\includegraphics[width=0.75\textwidth]{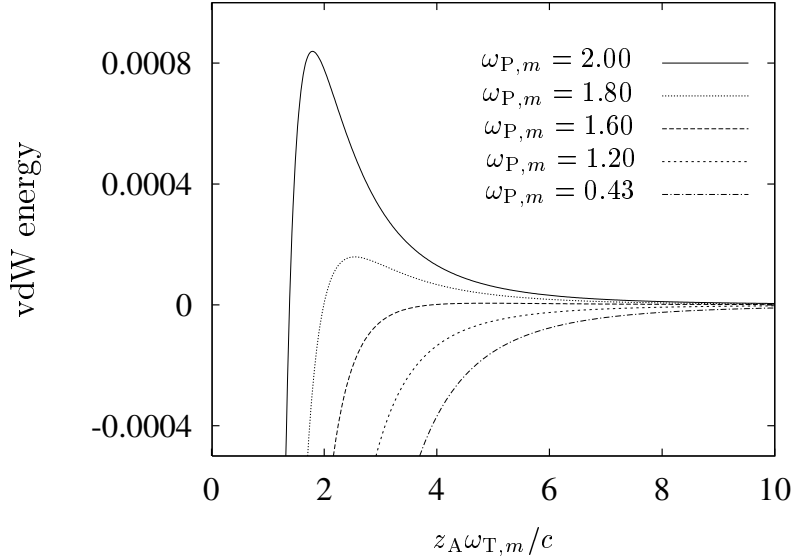}
\caption{
\label{Fig1}
The vdW energy $U_0(z_{\rm A})12\pi^2c/(\mu_0\omega_{{\rm T},m}^3d_{\rm A}^2)$ of a two-level atom in its ground state
situated above a semi-infinite magnetodielectric half-space as a function of the distance between the body and the
interface for different values of $\omega_{{\rm P},m}$ ($\omega_{{\rm P},e}/\omega_{{\rm T},m}$ $\!=$ $\!0.75$,
$\omega_{{\rm T},e}/\omega_{{\rm T},m}$ $\!=$ $\!1.03$,
$\gamma_e/\omega_{{\rm T},m}$ $\!=$ $\!\gamma_m/\omega_{{\rm T},m}$
$\!=$ $\!0.001$, $\omega_{10}/\omega_{{\rm T},m}$ $\!=$ $\!1$).
}
\end{center}
\end{figure}


\subsection{Ground-state atom in front of a magnetodielectric half-space}
\label{sec3.2}

To illustrate the influence of the magnetic properties on the CP force acting on an atom near a magnetodielectric body,
let us apply the theory to a two-level atom [$|0\rangle$, ground state; $|1\rangle$, excited state;
$\mathbf{d}_\mathrm{A}$ $\!\equiv$ $\!{\bf d}_{10}$ $\!=$ $\!d_{\rm A}(\cos\phi\sin\theta\,\mathbf{e}_x$ $\!+$
$\!\sin\phi\sin\theta\,\mathbf{e}_y$ $\!+$ $\!\cos\theta\,\mathbf{e}_z)$] situated above
($z_{\rm A}$ $\!>$ $\!0$) a semi-infinite magnetodielectric half-space ($z$ $\!<$ $\!0$). Using the appropriate
scattering Green tensor as given, e.g., in Ref.~\cite{Chew}, we find that
\begin{equation}
\label{eq37}
\bm{G}^{(1)}({\bf r}, {\bf r}, \omega)=\frac{i}{8\pi}\int_0^\infty {\rm d}q\,
\frac{q}{\beta_0}\,e^{2i\beta_0z}
\left\{\!r_s\!
\left(
\begin{array}{ccc}1&0&0\\ 0&1&0\\ 0&0&0\end{array}\right)
\!+r_p\,\frac{c^2}{\omega^2}\!
\left(
\begin{array}{rrr}
-\beta_0^2&0&0\\ 0&-\beta_0^2&0\\ 0&0&2q^2
\end{array}
\right)
\!\right\}\!,
\end{equation}
where
\begin{equation}
\label{eq38}
r_s=\frac{\mu\beta_0-\beta}{\mu\beta_0+\beta},\qquad
r_p=\frac{\varepsilon\beta_0-\beta}{\varepsilon\beta_0+\beta}
\end{equation}
denote the reflection coefficients for $s$- and $p$-polarized waves, respectively 
($\beta_0^2$ $\!=$ $\!\omega^2/c^2$ $\!-$ $\!q^2$ with ${\rm Im}\,\beta_0$ $\!>$ $\!0$,
$\beta^2$ $\!=$ $\!\varepsilon\mu\omega^2/c^2$ $\!-$ $\!q^2$ with ${\rm Im}\,\beta$ $\!>$ $\!0$).
Substituting this expression into Eq.~(\ref{eq35}) and recalling Eq.~(\ref{eq34}) [or equivalently using
Eqs.~(\ref{eq33}) and (\ref{eq34}) and averaging over all possible orientations of $\mathbf{d}_\mathrm{A}$],
we derive the following formula for the ground-state vdW energy ($\beta_0$ $\!=$ $\!ib_0$);
\begin{equation}
\label{eq39}
U_0(z_{\rm A})=\frac{\mu_0\omega_{10}d_{\rm A}^2}{12\pi^2}
\int_0^{\infty} \frac{{\rm d} u
\,u^2}{\omega_{10}^2+u^2}
\int_0^{\infty}\frac{{\rm d} q\,q}{b_0}e^{-2b_0z_{\rm A}}
\left\{\!r_s\!-r_p\,\left[2\left(\frac{cq}{\omega}\!\right)^2
+1\right]\right\}\!.
\end{equation}

Based upon a single-resonance permittivity and a single-resonance permeability of Drude-Lorentz type,
\begin{equation}
\label{eq40}
\varepsilon(\omega)=1
+\frac{\omega_{{\rm P},e}^2}{\omega_{{\rm T},e}^2-\omega^2-i\gamma_e\omega}\,,\qquad
\mu(\omega)=1
+\frac{\omega_{{\rm P},m}^2}{\omega_{{\rm T},m}^2-\omega^2-i\gamma_m\omega}\,,\qquad
\end{equation}
Fig.~\ref{Fig1} diplays $U_0(z_\mathrm{A})$ as function of $z_\mathrm{A}$ for different values of the magnetic plasma
frequency $\omega_{{\rm P},m}$. Note that while for short distances the dominant influence of the dielectric properties
always leads to the familiar attractive $1/z_{\rm A}^3$-potential (cf. Sec.~\ref{sec4.3}), strong magnetic properties
can give rise a repulsive potential barrier at some intermediate distances.


\section{The Casimir-Polder force}
\label{sec4}

Due to the presence of magnetodielectric bodies the structure of the electromagnetic vacuum can drastically change,
resulting in a body-induced shifting and broadening of atomic transition lines as the atom comes close the bodies. The
perturbative results in Sec.~\ref{sec3} obviously fail to take such effects into account. Further, the enhanced
spontaneous decay of an excited atom \cite{Ho01} leads to dynamical effects that cannot be described within the frame of
time-independent perturbation theory. In this section, we therefore use a quite distinct, non-perturbative approach to
the problem, by basing the calculation of the CP force on the Lorentz force that governs the atomic center-of-mass
motion.

\subsection{General result}
\label{sec4.1}

Using the multipolar Hamiltonian (\ref{eq1}) together with
Eqs.~(\ref{eq2}), (\ref{eq6}) and (\ref{eq8}), and recalling
the definitions (\ref{eq4}) and (\ref{eq10}) one can easily verify
\begin{equation}
\label{eq47}
     m_{\rm A} \dot{\hat{{\bf r}}}_{\rm A}
     =\frac{i}{\hbar} \bigl[\hat{H},m_{\rm A} \hat{{\bf r}}_{\rm A}\bigr]
       =\hat{{\bf p}}_{\rm A}
       + \hat{\bf d}\times \hat{{\bf B}}(\hat{\bf r}_{\rm A}),
\end{equation}
leading to
\begin{equation}
\label{eq48}
      m_{\rm A} \ddot{\hat{{\bf r}}}_{\rm A} =
      \hat{\mathbf{F}}
      = \frac{i}{\hbar} \bigl[\hat{H},
      \hat{{\bf p}}_{\rm A}\bigr]
      +\frac{\rm d}{{\rm d}t}\bigl[\hat{\bf d}
      \times \hat{{\bf B}}(\hat{\bf r}_{\rm A})\bigr]
      =\biggl\{\bm{\nabla}
      \bigl[\hat{{\bf d}} \hat{{\bf E}} ({\bf r})\bigr]
      +\frac{\rm d}{{\rm d}t}
      \bigl[\hat{{\bf d}} \times \hat{{\bf B}}({\bf r})\bigr]
      \biggr\}_{{\bf r}=\hat{{\bf r}}_{\rm A}},
\end{equation}
where in the last step magnetic dipole terms have been dropped in consistency with the electric dipole approximation
made and terms of the order of $O(v/c)$ (v, speed of the atomic center of mass) have been omitted in accordance with the
non-relativistic Hamiltonian (\ref{eq1}). Since $\hat{{\bf E}} ({\bf r})$ and $\hat{{\bf B}}({\bf r})$
are defined according to Eqs.~(\ref{eq11})--(\ref{eq13}), Eq.~(\ref{eq48}) describes the Lorentz force acting on an
atomic system in the presence of absorbing and dispersing magnetodielectric bodies and thus generalizes the well-known
free-space result \cite{Baxter93Lembessis93}.

Next, we determine the temporal evolution of the medium-assisted electromagnetic field with the aid of Hamiltonian
(\ref{eq1}) together with Eqs.~(\ref{eq2}), (\ref{eq6}), and (\ref{eq8}). Recalling the definitions (\ref{eq11}) and
(\ref{eq13}) and exploiting the commutation relations (\ref{eq7}), we may write
\begin{gather}
\label{eq49}
\underline{\hat{{\bf E}}}({\bf r},\omega,t)
= \underline{\hat{{\bf E}}}_{\rm free}({\bf r},\omega,t)
+ \underline{\hat{{\bf E}}}_{\rm source}({\bf r},\omega,t),\\
\label{eq50}
      \underline{\hat{{\bf E}}}_{\rm free}({\bf r},\omega,t)
      = \underline{\hat{{\bf E}}}({\bf r},\omega)e^{-i\omega t},\\
\label{eq51}
      \underline{\hat{{\bf E}}}_{\rm source}({\bf r},\omega,t)
      = \frac{i\mu_0}{\pi} \omega^2
      \!\int_0^t\!\! {\rm d}t'\,
      e^{-i\omega(t-t')}{\rm Im}\,\bm{G}[{\bf r},
      \hat{{\bf r}}_{\rm A}(t'),\omega]
      \hat{\bf d}(t'),
\end{gather}
where we have used similar approximations as in Eq.~(\ref{eq48}). In accordance with Eqs.~(\ref{eq11})--(\ref{eq13}),
we substitute this result into Eq.~(\ref{eq48}) and take the expectation with respect to the field state and the
internal (electronic) state of the atomic system, leading to
\begin{equation}
\label{eq52}
\bigl\langle\hat{\bf F}(t)\bigr\rangle
= \bigl\langle\hat{\bf F}_{\rm free}(t)\bigr\rangle
+ \bigl\langle\hat{\bf F}_{\rm source}(t)\bigr\rangle,
\end{equation}
\begin{eqnarray}
\label{eq53}
\lefteqn{
\bigl\langle\hat{\bf F}_{\rm free}(t)\bigr\rangle
 =\bigg\{\int_0^\infty {\rm d}\omega\, \bm{\nabla}
      \bigl\langle\hat{{\bf d}}(t)
      \underline{\hat{{\bf E}}}_{\rm free} ({\bf r},\omega,t)\bigr\rangle
}
\nonumber\\&&\hspace{8ex}
+\,\frac{1}{i\omega}\,\frac{\rm d}{{\rm d}t}
       \bigl\langle\hat{{\bf d}}(t)
       \times\! \big[ \bm{\nabla} \times
	\underline{\hat{{\bf E}}}_{\rm free} ({\bf r},\omega,t)\big]\bigr\rangle
      \bigg\}
      _{{\bf r}=\hat{{\bf r}}_{\rm A}(t)}
      +{\rm H.c.},
\end{eqnarray}
\begin{equation}
\label{eq54}
\bigl\langle\hat{\bf F}_{\rm source}(t)\bigr\rangle=
\bigl\langle\hat{\bf F}_{\rm source}^{\rm el}(t)\bigr\rangle
+\bigl\langle\hat{\bf F}_{\rm source}^{\rm mag}(t)\bigr\rangle,
\end{equation}
\begin{eqnarray}
\label{eq55}
\lefteqn{
\bigl\langle\hat{\bf F}_{\rm source}^{\rm el}(t)\bigr\rangle
= \biggl\{\frac{i\mu_0}{\pi}
\int_0^\infty {\rm d}\omega\,\omega^2
}
\nonumber\\&&\times\,
 \int_0^t\! {\rm d}t'\,
      e^{-i\omega(t-t')}
      \bm{\nabla}
      \bigl\langle\hat{\bf d}(t)
      {\rm Im}\,\bm{G} [{\bf r},\hat{{\bf r}}_{\rm A}(t'),\omega]
      \hat{\bf d}(t')\bigr\rangle
      \biggr\}
      _{{\bf r}=\hat{{\bf r}}_{\rm A}(t)}
      +{\rm H.c.},
\end{eqnarray}
\begin{eqnarray}
\label{eq56}
\lefteqn{
\bigl\langle\hat{\bf F}_{\rm source}^{\rm mag}(t)\bigr\rangle
= \biggl\{\frac{\mu_0}{\pi}\int_0^\infty {\rm d}\omega\,\omega
  \,\frac{\rm d}{{\rm d}t}\int_0^t\! {\rm d}t'\,
      e^{-i\omega(t-t')}
}
\nonumber\\&&\times\,
      \bigl\langle\hat{\bf d}(t)
      \times\,\bigl\{
      \bm{\nabla}\!\times
   {\rm Im}\,\bm{G} [{\bf r},\hat{{\bf r}}_{\rm A}(t'),\omega]
      \bigr\}
      \hat{\bf d}(t')\bigr\rangle
      \biggr\}
      _{{\bf r}=\hat{{\bf r}}_{\rm A}(t)}
      +{\rm H.c.}.
\end{eqnarray}
Equations (\ref{eq52})--(\ref{eq56}) apply to both driven and non-driven atomic systems and to both weak and strong
atom-field coupling. In particular, when the medium-assisted electromagnetic field is initially in the ground state,
then $\bigl\langle\hat{\bf F}_{\rm free}(t)\bigr\rangle$ $\!=$ $\!0$ is valid and Eqs.~(\ref{eq55}) and (\ref{eq56}),
respectively, just determine the electric part and the magnetic part of the time-dependent CP force acting on an atomic
system prepared in an arbitrary internal quantum state.


\subsection{Weak-coupling regime}
\label{sec4.2}

Representing the electric dipole operator in the form
\begin{equation}
\label{eq57}
      \hat{\bf d}(t)
     = \sum_{m,n}
     \mathbf{d}_{mn} \hat{A}_{mn}(t)
\end{equation}
($\hat{A}_{mn}$ $\!=$ $\!|m\rangle\langle n|$, with $|n\rangle,\,|m\rangle$ being the internal atomic energy
eigenstates), we may express the dipole-dipole correlation function appearing in Eqs.~(\ref{eq55}) and (\ref{eq56})
in terms of correlation functions of the $\hat{A}_{mn}(t)$ as
\begin{equation}
\label{eq58}
\bigl\langle\hat{\bf d}(t)\otimes\hat{\bf d}(t')\bigr\rangle
= \sum_{m,n}\sum_{m',n'} {\bf d}_{mn}\otimes {\bf d}_{m'n'}
\bigl\langle\hat{A}_{mn}(t)\hat{A}_{m'n'}(t')\bigr\rangle.
\end{equation}
In the weak-coupling regime, the Markov approximation can be exploited, and the correlation functions 
$\langle\hat{A}_{mn}(t)\hat{A}_{m'n'}(t')\rangle$ can be calculated by means of the quantum regression theorem (see,
e.g., Ref.~\cite{Vogel01}). For a non-degenerate atomic system with sufficiently slow center-of-mass motion this leads
to (App.~\ref{AppB})
\begin{equation}
\label{eq59}
       \bigl\langle \hat{A}_{mn}(t)
       \hat{A}_{m'n'}(t')\bigr\rangle
       = \delta_{nm'}
       \bigl\langle \hat{A}_{mn'}(t')\bigr\rangle
        e^{\{i\tilde{\omega}_{mn}(\hat{\mathbf{r}}_\mathrm{A})
       -[\Gamma_m(\hat{\mathbf{r}}_\mathrm{A})
       +\Gamma_n(\hat{\mathbf{r}}_\mathrm{A})]/2\}(t-t')}
\end{equation}
[$t$ $\!\ge$ $\!t'$, $m$ $\!\neq$ $n$, $\hat{\mathbf{r}}_\mathrm{A}$ $\!=$ $\!\hat{\mathbf{r}}_\mathrm{A}(t)$], where
\begin{gather}
\label{eq60}
         \tilde{\omega}_{mn}(\hat{\mathbf{r}}_\mathrm{A})
         =
         \omega_{mn}
         +\delta\omega_m(\hat{\mathbf{r}}_\mathrm{A})
         -\delta\omega_n(\hat{\mathbf{r}}_\mathrm{A}),\\
\label{eq61}
   \delta\omega_m(\hat{\mathbf{r}}_\mathrm{A})
   =\sum_k \delta\omega_m^k(\hat{\mathbf{r}}_\mathrm{A}),\\
\label{eq62}
        \delta\omega_m^k(\hat{\mathbf{r}}_\mathrm{A})
         = \frac{\mu_0}{\pi\hbar}
	{\cal P}\int_0^\infty \!\!{\rm d}\omega\, \omega^2
        \frac{{\bf d}_{km}
	{\rm Im}\bm{G}^{(1)}\,
	(\hat{{\bf r}}_{\rm A},\hat{{\bf r}}_{\rm A},\omega)
	{\bf d}_{mk}}{\tilde{\omega}_{mk}(\hat{{\bf r}}_{\rm A})-\omega}
\end{gather}
are the body-induced shifted transition frequencies, and
\begin{gather}
\label{eq63}
         \Gamma_m(\hat{\mathbf{r}}_\mathrm{A})
         = \sum_k \Gamma_m^k(\hat{\mathbf{r}}_\mathrm{A}),\\
\label{eq64}
        \Gamma_m^k(\hat{\mathbf{r}}_\mathrm{A})
        = \frac{2\mu_0 }{\hbar}\,
	\Theta[\tilde{\omega}_{mk}(\hat{{\bf r}}_{\rm A})]
        [\tilde{\omega}_{mk}(\hat{{\bf r}}_{\rm A})]^2{\bf d}_{km}
	{\rm Im}\bm{G}\,
	[\hat{{\bf r}}_{\rm A},\hat{{\bf r}}_{\rm A},
        \tilde{\omega}_{mk}(\hat{{\bf r}}_{\rm A})]
	{\bf d}_{mk}
\end{gather}
are the position-dependent level widths. Note that the position-in\-de\-pen\-dent (infinite) Lamb-shift terms resulting
from $\bm{G}^{(0)}\,(\hat{{\bf r}}_{\rm A},\hat{{\bf r}}_{\rm A},\omega)$ [recall Eq.~(\ref{eq28})] have been absorbed
in the transition frequencies $\omega_{mn}$. Equation (\ref{eq62}) can be rewritten by changing to imaginary
frequencies (cf. Sec.~\ref{sec3.1}), resulting in
\begin{eqnarray}
\label{eq65}
        \delta\omega_m^k(\hat{\mathbf{r}}_\mathrm{A})&=&
        -\frac{\mu_0}{\hbar}\Theta[\tilde{\omega}_{mk}(\hat{{\bf r}}_{\rm A})]
      [\tilde{\omega}_{mk}(\hat{{\bf r}}_{\rm A})]^2
      {\bf d}_{km}{\rm Re}\,\bm{G}^{(1)}\,
	[\hat{{\bf r}}_{\rm A},\hat{{\bf r}}_{\rm A},
        \tilde{\omega}_{mk}(\hat{{\bf r}}_{\rm A})]
	{\bf d}_{mk}
        \nonumber\\
        &&+\frac{\mu_0}{\pi\hbar}
	\int_0^\infty {\rm d}u\, u^2\tilde{\omega}_{km}(\hat{{\bf r}}_{\rm A})
        \frac{{\bf d}_{km}
	\bm{G}^{(1)}
	(\hat{{\bf r}}_{\rm A},\hat{{\bf r}}_{\rm A},iu)
	{\bf d}_{mk}}{[\tilde{\omega}_{km}(\hat{{\bf r}}_{\rm A})]^2+u^2}\,.
\end{eqnarray}
The calculation of
\begin{equation}
\label{eq66}
\big\langle\hat{A}_{mn}(t)\big\rangle
=\sigma_{nm}(t)
\end{equation}
[$\sigma_{nm}(0)$ $\!=$ $\!\sigma_{nm}$] yields (App.~\ref{AppB})
\begin{equation}
\label{eq67}
       \sigma_{nm}(t)
       = e^{\{i\tilde{\omega}_{mn}(\hat{{\bf r}}_{\rm A})
       -[\Gamma_m(\hat{{\bf r}}_{\rm A})+\Gamma_n(\hat{{\bf r}}_{\rm A})]/2\}t}
       \sigma_{nm}
\end{equation}
for $m$ $\!\neq$ $\!n$, so the remaining task consists in solving the balance equations
\begin{equation}
\label{eq68}
       \dot{\sigma}_{mm}(t)
       = -\Gamma_m(\hat{{\bf r}}_{\rm A})
       \sigma_{mm}(t)
       + \sum_n \Gamma_n^m(\hat{{\bf r}}_{\rm A})
              \sigma_{nn}(t).
\end{equation}

Upon using Eqs.~(\ref{eq58}), (\ref{eq59}), and (\ref{eq66}), the CP force as defined according to
Eqs.~(\ref{eq54})--(\ref{eq56}) can now be calculated by evaluating the time integrals in the spirit of the Markov
approximation, resulting in
\begin{gather}
\label{eq69}
\bigl\langle{\hat{\bf F}}(t)\bigr\rangle
=\sum_{m,n} \sigma_{nm}(t){{\bf F}}_{mn}(\hat{\bf r}_{\rm A}),\\
\label{eq70}
{{\bf F}}_{mn}(\hat{\bf r}_{\rm A})
={{\bf F}}_{mn}^{\rm el}(\hat{\bf r}_{\rm A})
+{{\bf F}}_{mn}^{\rm mag}(\hat{\bf r}_{\rm A}),
\end{gather}
with
\begin{gather}
\label{eq71}
      {{\bf F}}_{mn}^{\rm el}(\hat{\bf r}_{\rm A}) =
      \Biggl\{
      \frac{\mu_0}{\pi} \sum_{k}
      \int_0^\infty {\rm d}\omega\,
      \omega^2
      \frac{\bm{\nabla}\otimes
      {\bf d}_{mk}
      {\rm Im}\,\bm{G}^{(1)} ({\bf r},
      \hat{{\bf r}}_{\rm A},\omega) {\bf d}_{kn} }
      {\omega+\tilde{\omega}_{kn}(\hat{\mathbf{r}}_\mathrm{A})
      -i[\Gamma_k(\hat{\mathbf{r}}_\mathrm{A})
      +\Gamma_m(\hat{\mathbf{r}}_\mathrm{A})]/2}
      \Biggr\}_{{\bf r}=\hat{{\bf r}}_{\rm A}}
      \!\!+\, {\rm H.c.},\\
\label{eq72}
     {{\bf F}}^{\rm mag}_{mn}(\hat{\bf r}_{\rm A})=
     \Biggl\{\frac{\mu_0}{\pi}\sum_k
     \int_0^\infty {\rm d}\omega\,\omega
     \tilde{\omega}_{mn}(\hat{\mathbf{r}}_\mathrm{A})
     \frac{{\bf d}_{mk}\times\bigl[\bm{\nabla}\times
     {\rm Im}\,\bm{G}^{(1)} ({\bf r},\hat{{\bf r}}_{\rm A},\omega)\bigr]
     {\bf d}_{kn}}{\omega+\tilde{\omega}_{kn}(\hat{\mathbf{r}}_\mathrm{A})
     -i[\Gamma_k(\hat{\mathbf{r}}_\mathrm{A})
     +\Gamma_m(\hat{\mathbf{r}}_\mathrm{A})]/2}
     \Biggr\}_{{\bf r}=\hat{{\bf r}}_{\rm A}}
     \!\!+\,{\rm H.c.},
\end{gather}
where we have again made the replacement $\bm{G}({\bf r},\hat{\bf r}_{\rm A},\omega)\mapsto
\bm{G}^{(1)}({\bf r}, \hat{\bf r}_{\rm A},\omega)$ [cf. Eq.~(\ref{eq28})]. As $\hat{{\bf r}}_{\rm A}$ effectively
enters equations (\ref{eq69})--(\ref{eq72}) as a parameter, the caret will be removed in the following
($\hat{\mathbf{r}}_\mathrm{A}$ $\!\mapsto$ $\!\mathbf{r}_\mathrm{A}$). We finally rewrite Eqs.~(\ref{eq71}) and
(\ref{eq72}) by going over to imaginary frequencies (cf. Sec.~\ref{sec3.1}). Introducing the abbreviating notation
\begin{equation}
\label{eq73}
\Omega_{mnk}(\mathbf{r}_\mathrm{A})
=\tilde{\omega}_{nk}(\mathbf{r}_\mathrm{A})
+i[\Gamma_m(\mathbf{r}_\mathrm{A})
+\Gamma_k(\mathbf{r}_\mathrm{A})]/2
\end{equation}
as well as the generalized atomic polarizability tensor
\begin{eqnarray}
\label{eq74}
\lefteqn{
      \bm{\alpha}_{mn}(\mathbf{r}_\mathrm{A},\omega) =
      \frac{1}{\hbar} \sum_k
       \biggl[
        \frac{{\bf d}_{mk} \otimes {\bf d}_{kn}}
       {\tilde{\omega}_{kn}(\mathbf{r}_\mathrm{A}) - \omega
       - i(\Gamma_k[\mathbf{r}_\mathrm{A})
       +\Gamma_m(\mathbf{r}_\mathrm{A})]/2}
}
\nonumber\\&&\hspace{10ex}
       +\,\frac{{\bf d}_{kn} \otimes {\bf d}_{mk}}
       {\tilde{\omega}_{km}(\mathbf{r}_\mathrm{A}) + \omega
       + i[\Gamma_k(\mathbf{r}_\mathrm{A})+\Gamma_n(\mathbf{r}_\mathrm{A})]/2}
       \biggr],
\end{eqnarray}
with $\bm{\alpha}_l(\mathbf{r}_\mathrm{A},\omega)$ $\!=$ $\!\bm{\alpha}_{ll}(\mathbf{r}_\mathrm{A},\omega)$ being the
ordinary (Kramers-Kronig-con\-sist\-ent) polarizability tensor of an atom in state $|l\rangle$ (cf.
Ref.~\cite{Fain63Milonni04}), we derive
\begin{gather}
\label{eq75}
{{\bf F}}_{mn}^{\rm el}({\bf r}_{\rm A})
={\bf F}_{mn}^{\rm el, or}({\bf r}_{\rm A})
+{\bf F}_{mn}^{\rm el, r}({\bf r}_{\rm A}),\\
\label{eq76}
{{\bf F}}_{mn}^{\rm mag}({\bf r}_{\rm A})
={\bf F}_{mn}^{\rm mag, or}({\bf r}_{\rm A})
+{\bf F}_{mn}^{\rm mag, r}({\bf r}_{\rm A}),
\end{gather}
where
\begin{gather}
\label{eq77}
     {{\bf F}}_{mn}^{\rm el, or}({\bf r}_{\rm A})=
     -\biggl\{
     \frac{\hbar\mu_0}{2\pi}
     \int_0^\infty {\rm d}u\, u^2
     \bigl[(\alpha_{mn})_{ij}(\mathbf{r}_\mathrm{A},iu)
     +(\alpha_{mn})_{ij}(\mathbf{r}_\mathrm{A},-iu)\bigr]
     \bm{\nabla}G^{(1)}_{ij}({\bf r},{\bf r}_{\rm A},iu)
     \biggr\}_{{\bf r}={\bf r}_{\rm A}},
\\
\label{eq78}
{{\bf F}}_{mn}^{\rm el,r}({\bf r}_{\rm A})=
      \biggl\{\mu_0\sum_{k}
     \Theta({\tilde{\omega}_{nk}})
     \Omega^2_{mnk}(\mathbf{r}_\mathrm{A})
     \bm{\nabla}\otimes{\bf d}_{mk}\bm{G}^{(1)}[{\bf r},{\bf r}_{\rm A},
     \Omega_{mnk}(\mathbf{r}_\mathrm{A})]
     {\bf d}_{kn}
     \biggr\}_{{\bf r}={\bf r}_{\rm A}}+{\rm H.c.},
\end{gather}
\begin{gather}
\label{eq79}
\begin{split}
     {{\bf F}}_{mn}^{\rm mag, or}({\bf r}_{\rm A})
     =\,&\biggl\{
     \frac{\hbar\mu_0}{2\pi}
      \int_0^\infty {\rm d}u\, u^2 \,{\rm Tr}\,
      \biggl(\biggl[
      \frac{\tilde{\omega}_{mn}(\mathbf{r}_\mathrm{A})}{iu}\,
      \bm{\alpha}_{mn}^\top(\mathbf{r}_\mathrm{A},iu)
     \\&
      - \frac{\tilde{\omega}_{mn}(\mathbf{r}_\mathrm{A})}{iu}\,
      \bm{\alpha}_{mn}^\top(\mathbf{r}_\mathrm{A},-iu)
      \biggr]\times
      \big[\bm{\nabla}\times\bm{G}^{(1)}({\bf r},{\bf r}_{\rm A},iu)\big]
      \biggr)\biggr\}_{{\bf r}={\bf r}_{\rm A}},
\end{split}
\\
\label{eq80}
\begin{split}
    {{\bf F}}_{mn}^{\rm mag,r}({\bf r}_{\rm A}) =
     \,&\biggl\{\mu_0\sum_{k}
     \Theta({\tilde{\omega}_{nk}})
     \tilde{\omega}_{mn}(\mathbf{r}_\mathrm{A})
     \Omega_{mnk}(\mathbf{r}_\mathrm{A})
     {\bf d}_{mk}
\\&\times\,
     \left(\bm{\nabla}
     \times\bm{G}^{(1)}[{\bf r},{\bf r}_{\rm A},
     \Omega_{mnk}(\mathbf{r}_\mathrm{A})]
     {\bf d}_{kn}
     \right)
     \biggr\}_{{\bf r}={\bf r}_{\rm A}}
     +{\rm H.c.}
\end{split}
\end{gather}
[$({\rm Tr}\,\bm{T})_j$ $\!=$ $\!T_{iji}$].

Equation (\ref{eq69}) together with Eq.~(\ref{eq70}) and Eqs.~(\ref{eq75})--(\ref{eq80}) is the natural generalization
of the perturbative result of Sec.~\ref{sec3.1}. In particular, an atom intially prepared in an energy eigenstate
$|l\rangle$ experiences for $\Gamma_C t$ $\!\ll$ $\!1$ ($\Gamma_C$, characteristic atomic decay rate) the CP force
$\bigl\langle{\hat{\bf F}}(t)\bigr\rangle$
$\!\simeq$ $\!\bigl\langle{\hat{\bf F}}(0)\bigr\rangle$
$\!=$ $\!{\mathbf{F}}_{ll}^\mathrm{el}(\mathbf{r}_\mathrm{A})$
$\!=$ $\!{\mathbf{F}}_{ll}^\mathrm{el, or}(\mathbf{r}_\mathrm{A})$
$\!+$ $\!{\mathbf{F}}_{ll}^\mathrm{el, r}(\mathbf{r}_\mathrm{A})$
with
\begin{gather}
\label{eq81}
{\mathbf{F}}_{ll}^\mathrm{el, or}(\mathbf{r}_\mathrm{A})
     =     -\frac{\hbar\mu_0}{4\pi}
      \int_0^\infty {\rm d}u u^2\big[(\alpha_l)_{ij}
      ({\bf r}_{\rm A},iu)
      +(\alpha_l)_{ij}({\bf r}_{\rm A},-iu)\big]
     \bm{\nabla}_{\!\!{\rm A}}G^{(1)}_{ij}
     ({\bf r}_{\rm A},{\bf r}_{\rm A},iu),\\
\label{eq82}
{\mathbf{F}}_{ll}^\mathrm{el,r}(\mathbf{r}_\mathrm{A})
     =\frac{\mu_0}{2}\sum_{k}
      \Theta({\tilde{\omega}_{lk}})
     \Omega_{lk}^2({\bf r}_{\rm A})
     \left\{
     \bm{\nabla}
     \otimes{\bf d}_{lk}\bm{G}^{(1)}[{\bf r},{\bf r},
     \Omega_{lk}({\bf r}_{\rm A})]{\bf d}_{kl}
     \right\}_{\mathbf{r}=\mathbf{r}_\mathrm{A}}
     \!\!+{\rm H.c.}
\end{gather}
[$\Omega_{lk}({\bf r}_{\rm A})$ $\!\equiv$ $\!\Omega_{llk}({\bf r}_{\rm A})$; ${\bf d}_{lk}$ real].
Note that due to the position dependence of the atomic polarizability even the ground-state CP-force
${\mathbf{F}}_{00}^\mathrm{el}(\mathbf{r}_\mathrm{A})$ $\!=$
$\!{\mathbf{F}}_{00}^\mathrm{el,or}(\mathbf{r}_\mathrm{A})$ cannot be derived from a potential in the usual way.
It is worth noting that when the atom is initially prepared in a coherent superposition of states, then the
corresponding off-diagonal force components $\sigma_{nm}(t)\mathbf{F}_{mn}(\mathbf{r}_\mathrm{A})$
($n$ $\!\neq$ $\!m$) can also contribute to the total CP force as given by Eq.~(\ref{eq69}). These components
contain contributions not only from the electric part of the Lorentz force but also from the magnetic part, as can be
easily seen from inspection of Eqs.~(\ref{eq79}) and (\ref{eq80}).


\subsection{Excited atom in front of a magnetodielectric half-space}
\label{sec4.3}

To illustrate the effects of body-induced level shifting and broadening, let us again consider a two-level atom in front
of a magnetodielectric half-space (cf. Sec.~\ref{sec3.2}) and calculate the CP force ${\mathbf{F}}_{11}(z_\mathrm{A})$
acting on the atom in the upper state. For simplicity, we restrict our attention to the short-distance limit $z_{\rm A}
\sqrt{|\varepsilon\mu|}|\omega|/c$ $\!\ll$ $\!1$, where the approximations $\beta_0$ $\!\simeq$ $\!iq$, $\beta$
$\!\simeq$ $\!iq$ can be applied to Eq.~(\ref{eq37}). In this limit, which is governed by the dielectric properties
of the material, Eq.~(\ref{eq60}) [together with Eqs.~(\ref{eq61}) and (\ref{eq65})] and Eq.~(\ref{eq63}) [together with
Eq.~(\ref{eq64})] approximate to
\begin{equation}
\label{eq83}
\tilde{\omega}_{10}(z_\mathrm{A})
= \omega_{10}+\delta\omega_1(z_\mathrm{A})-\delta\omega_0(z_\mathrm{A})
= \omega_{10}- \frac{d_{\rm A}^2(1+\cos^2\theta)}{32\pi\varepsilon_0\hbar z_{\rm A}^3}
\frac{|\varepsilon[\tilde{\omega}_{10}(z_\mathrm{A})]|^2-1}
{|\varepsilon[\tilde{\omega}_{10}(z_\mathrm{A})]+1|^2}\,,
\end{equation}
\begin{equation}
\label{eq84}
\Gamma_1(z_\mathrm{A})=
\frac{d_{\rm A}^2(1+\cos^2\theta)}{8\pi\varepsilon_0\hbar z_{\rm A}^3}
\frac{{\rm Im}\,\varepsilon[\tilde{\omega}_{10}(z_\mathrm{A})]}
{|\varepsilon[\tilde{\omega}_{10}(z_\mathrm{A})]+1|^2}\,,
\end{equation}
where we have neglected the off-resonant second term in Eq.~(\ref{eq65}) as well as the small free-space decay rate.
Taking into account that ${\mathbf{F}}_{11}(z_\mathrm{A})$ $\!=$ $\!{\mathbf{F}}_{ll}^\mathrm{el}(z_\mathrm{A})$
$\!\simeq$ $\!{\mathbf{F}}_{ll}^\mathrm{el, r}(z_\mathrm{A})$ is valid, from Eq.~(\ref{eq82}) [together with
Eq.~(\ref{eq37}) (in the short-distance limit) and Eqs.~(\ref{eq83}) and (\ref{eq84})] we finally obtain
\begin{equation}
\label{eq85}
{\mathbf{F}}_{11}^\mathrm{r}(z_\mathrm{A})
=F_{11}^\mathrm{r}(z_\mathrm{A})\,\mathbf{e}_z
=-\frac{3d_{\rm A}^2(1+\cos^2\theta)}{32\pi\varepsilon_0z_{\rm A}^4}
\frac{|\varepsilon[\Omega_{10}(z_\mathrm{A})]|^2-1}
{|\varepsilon[\Omega_{10}(z_\mathrm{A})]+1|^2}\,\mathbf{e}_z,
\end{equation}
where, according to Eq.~(\ref{eq73}), $\Omega_{10}(z_\mathrm{A})$ $\!=$
$\!\tilde{\omega}_{10}(z_\mathrm{A})+i\Gamma_1(z_\mathrm{A})/2$.

\begin{figure}[htb]
\noindent

\psfrag{CP force}{\large CP force}
\psfrag{w/wTe}{$\omega_{10}/\omega_{{\rm T},e}$}

\noindent
\begin{center}
\includegraphics[width=0.75\textwidth]{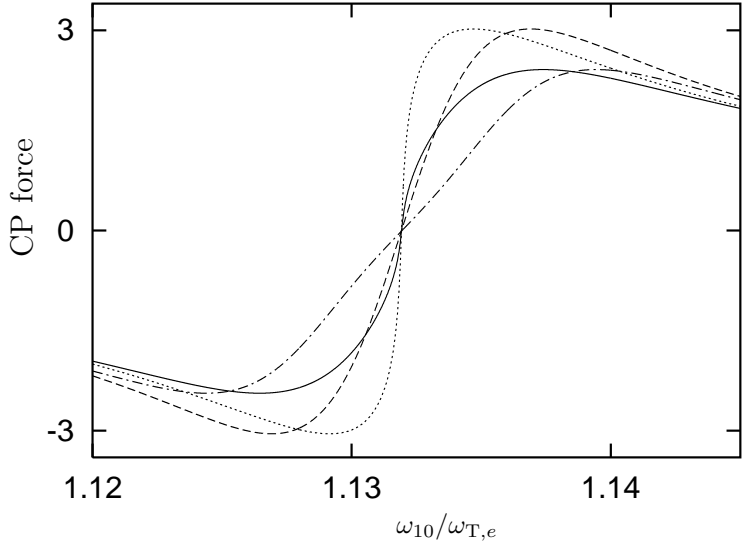}
\end{center}
\caption{
\label{Fig2}
The resonant part of the CP force $F_{11}^\mathrm{r}(z_{\rm A})\lambda_{{\rm T},e}^4
16\pi\varepsilon_0\times 10^{-9}/(3d_{\rm A}^2)$ on a two-level atom situated at distance 
$z_\mathrm{A}/\lambda_{\mathrm{T},e}$ $\!=$ $\!0.0075$ of a semi-infinite half-space medium with transition dipole 
moment perpendicular to the interface as a function of bare frequency (solid lines), where we have chosen the following
parameters; $\omega_{{\rm P},e}/\omega_{{\rm T},e}$ $\!=$ $\!0.75$, $\gamma_e/\omega_{{\rm T},e}$ $\!=$ $\!0.01$,
${\omega}_{{\rm T},e}^2d_{\rm A}^2/(3\pi\hbar\varepsilon_0c^3)$ $\!=$ $\!10^{-7}$.
For comparison, both the perturbative result (dashed lines)
and the separate effects of level shifting (dotted lines) and
level broadening (dash-dotted lines) are shown.
}
\end{figure}

For a single-resonance medium according to Eq.~(\ref{eq40}), Fig.~\ref{Fig2} displays $F^\mathrm{r}_{11}(z_\mathrm{A})$
as a function of the bare transition frequency $\omega_{10}$. It is seen that the typical dispersion profile of
the attractive/repulsive CP force around the (surface-plasmon induced) resonance frequency \mbox{$\omega_{\rm S}$ $\!=$
$\!\sqrt{\omega_{{\rm T},e}^2+\omega_{{\rm P},e}^2/2}$}, which is already known from perturbation theory, is noticeably
reduced by the decay-induced damping while retaining its width, because the competing effects of decay-induced
broadening and narrowing due to the frequency shifts almost cancel.

\section{Summary}
\label{sec6}

Within the frame of electromagnetic-field quantization in the presence of dispersing and absorbing linear media, we
have presented a consistent theory for calculating the CP force experienced by an atomic system in the presence of
an arbitrary arrangement of magnetodielectric bodies. Extending previous results \cite{Buhmann03} to the case
of magnetodielectric media, we have derived the vdW energy of an atomic system prepared in an energy eigenstate
using perturbative methods and demonstrated that magnetic properties of the media can give rise to interesting
effects such as the formation of a potential barrier for ground-state atoms above a magnetodielectric half-space.
In an alternative approach based on the Lorentz force that governs the atomic center-of mass motion, we have derived an
expression for the CP force that applies to arbitrary atomic states and both strong and weak atom-field coupling.
Restricting our attention to a non-driven atom in the weak-coupling regime, we have found that the CP force can be
written as a superposition of force components weighted by the time-dependent intra-atomic density matrix elements that
solve the intra-atomic master equation. Each force component can be written in terms of the Green tensor for the
electromagnetic field and appropriate atomic quantities such as the polarizability. In contrast to the perturbative
result, the atomic quantities exhibit body-induced shifts and broadenings of the atomic transition lines which can
noticeably influence the CP force when the atomic system comes close to a body.


\vspace{2ex}
\noindent
{\bf Acknowledgement:}
S.Y.B. acknowledges valuable discussions with O. P. Sushkov as well as M.-P. Gorza. This work was supported by the
Deutsche Forschungsgemeinschaft and the SaxoSmithKline Stiftung. S.Y.B. is grateful for being granted a
Th\"{u}\-rin\-ger Landesgraduiertenstipendium.

\appendix

\section{Intra-atomic equations of motion}
\label{AppB}

Upon using the Hamiltonian (\ref{eq1}) together with Eqs.~(\ref{eq6}),
(\ref{eq8}), and (\ref{eq22}), and exploiting the
decomposition (\ref{eq57}) the following equations of
motion can be derived;
\begin{eqnarray}
\label{B1}
       \dot{\hat{A}}_{mn}
       = \frac{i}{\hbar}\big[\hat{H},\hat{A}_{mn}\big]
       &\hspace{-1ex}=&\hspace{-1ex}  i\omega_{mn} \hat{A}_{mn}
       +\frac{i}{\hbar} \!\sum_k
       \left[
       \big({\bf d}_{nk}\hat{A}_{mk}\!-\!{\bf d}_{km}\hat{A}_{kn}\big)
       \int_0^\infty\!\!  {\rm d}\omega\,
       \underline{\hat{\bf E}}(\hat{\bf r}_{\rm A},\omega)
       \right.\nonumber\\
       &&\left.
       +\!\int_0^\infty\!\! {\rm d}\omega\,
       \underline{\hat{\bf E}}{}^{\dagger}
       (\hat{\bf r}_{\rm A},\omega)
        \big({\bf d}_{nk}\hat{A}_{mk}
       \!-\!{\bf d}_{km}\hat{A}_{kn}\big)
       \right],
\quad
\end{eqnarray}
where a similar approximation as in Eq.~(\ref{eq48}) has been made. For weak atom-field coupling and
sufficiently slow center-of-mass motion, the source-field dynamics as given by Eqs.~(\ref{eq51}) can be approximated by
carrying out the time integral in the Markov approximation, resulting in
\begin{gather}
\label{B2}
\int_0^\infty \!\! {\rm d}\omega\,
\underline{\hat{{\bf E}}}_{\rm source}
(\hat{{\bf r}}_\mathrm{A},\omega)
= \sum_{m,n} {\bf g}_{mn}(\hat{{\bf r}}_\mathrm{A})
 \hat{A}_{mn},\\
\label{F3}
       {\bf g}_{mn}(\hat{{\bf r}}_\mathrm{A})
       =\frac{i\mu_0}{\pi}\int_0^\infty \!\!{\rm d}\omega\, \omega^2
        {\rm Im}\, \bm{G}(\hat{{\bf r}}_{\rm A},
	         \hat{{\bf r}}_{\rm A},\omega){\bf d}_{mn}
       \zeta[\tilde{\omega}_{nm}(\hat{{\bf r}}_{\rm A})\!-\!\omega]
\end{gather}
[$\zeta(x)$ $\!=$ $\!\pi\delta(x)$ $\!+$ $\!i{\cal P}/x$], with $\tilde{\omega}_{nm}(\hat{{\bf r}}_{\rm A})$
according to Eq.~(\ref{eq60}).

We substitute Eqs.~(\ref{eq49}), (\ref{eq50}), and (\ref{B2}) into Eq.~(\ref{B1}) and take expectation values with
respect to the internal atomic state and the vacuum field state, noting that in this case the contributions from the
free-field part (\ref{eq50}) vanish. Upon applying the decomposition
\begin{equation}
\label{B3}
   \frac{i}{\hbar}{\bf d}_{nk}{\bf g}_{kn}
   (\hat{\mathbf{r}}_\mathrm{A})
	= - i \delta\omega_n^k(\hat{\bf r}_{\rm A})
        - {\textstyle\frac{1}{2}} \Gamma_n^k(\hat{\bf r}_{\rm A}),
\end{equation}
and using the fact that for a non-degenerate atom off-diagonal density matrix elements decouple from each other as well
as from the diagonal ones, one can derive Eqs.~(\ref{eq59}), (\ref{eq67}), and (\ref{eq68}).



\begin{thebibliography}{99}

\bibitem{Chesters73}
M. A. Chesters, M. Hussain, and J. Pritchard, Surf. Sci. \textbf{35}, 161 (1973).

\bibitem{Binnig86}
G. Binnig, C. F. Quate, and C. Gerber, Phys. Rev. Lett. \textbf{56}, 930 (1986).

\bibitem{Shimizu02}
F. Shimizu and J. Fujita, Phys. Rev. Lett. \textbf{88}, 123201 (2002).

\bibitem{Casimir48}
H. B. G. Casimir and D. Polder, Phys. Rev. \textbf{73}, 360 (1948).

\bibitem{Milloni92}
P. W. Milonni and M.-L. Shih, Phys. Rev. A \textbf{45}, 4241 (1992).

\bibitem{Buhmann03}
S. Y. Buhmann, Ho Trung Dung, and D.-G. Welsch, J. Opt. B: Quantum Semiclass. Opt. \textbf{6}, 127 (2004).

\bibitem{McLachlan63}
A. D. McLachlan, Proc. R. Soc. London Ser. A \textbf{271}, 387 (1963);
Mol. Phys. \textbf{7}, 381 (1963).

\bibitem{WylieSipe}
J. M. Wylie and J. E. Sipe, Phys. Rev. A \textbf{30}, 1185 (1984); \textbf{32}, 2030 (1985).

\bibitem{Henkel02}
C. Henkel, K. Joulain, J.-P. Mulet, and J.-J. Greffet, J. Opt. A: Pure Appl. Opt. \textbf{4}, 109 (2002).

\bibitem{Craig84}
D. P. Craig and T. Thirunamachandran \textit{Molecular Quantum Electrodynamics} (Academic Press, New York, 1984).

\bibitem{Knoll01}
L. Kn\"{o}ll, S. Scheel, and D.-G. Welsch, in \textit{Coherence and Statistics of Photons and Atoms},
edited by J. Pe\v{r}ina (Wiley, New York, 2001), p. 1.

\bibitem{Ho03}
S. Y. Buhmann, Ho Trung Dung, J. K\"{a}stel, L. Kn\"{o}ll, S. Scheel, and D.-G. Welsch, Phys. Rev. A \textbf{68},
043816 (2003).

\bibitem{Veselago68}
V. G. Veselago, Sov. Phys. Usp. \textbf{10}, 509 (1968).

\bibitem{Chew}
W. C. Chew, \textit{Waves and Fields in Inhomogeneous Media} (IEEE Press, New York, 1995), Secs. 2.1.3, 2.1.4, and
7.4.2.

\bibitem{Ho01}
Ho Trung Dung, L. Kn\"{o}ll, and D.-G. Welsch, Phys. Rev. A \textbf{64}, 013804 (2001).

\bibitem{Baxter93Lembessis93}
C. Baxter, M. Babiker, and R. Loudon, Phys. Rev. A \textbf{47}, 1278 (1993); V. E. Lembessis, M. Babiker, C. Baxter, and
R. Loudon, \textit{ibid.} \textbf{48}, 1594 (1993).

\bibitem{Vogel01}
W. Vogel, D.-G. Welsch, and S. Wallentowitz, \textit{Quantum Optics, An Introduction} (Wiley-VCH, Berlin, 2001).

\bibitem{Fain63Milonni04}
V. M. Fain and Y. I. Khanin, \textit{Quantum Electronics} (Cambridge, Mass., MIT Press, 1969);
P. W. Milonni and R. W. Boyd, Phys. Rev. A \textbf{69}, 023814 (2004).

\end{thebibliography}
\end{document}